\documentclass[trackchanges,twocolumn]{aastex7}

\usepackage{CJK}
\usepackage{framed,amsmath,bm,booktabs}
\usepackage{float}

\graphicspath{{./}{figures}}

\submitjournal{ApJ}
\shorttitle{Observing Patchy Reionization with CMB Optical-Depth Fluctuations}
\shortauthors{Takoudes, Zhu, and Gnedin}

\begin{document}
\begin{CJK*}{UTF8}{gkai}

\title{The Challenge of Observing Patchy Reionization with CMB Optical-Depth Fluctuations}

\correspondingauthor{Hanjue Zhu (朱涵珏)}

\author{Nick Takoudes}
\affiliation{Department of Astronomy \& Astrophysics; 
The University of Chicago; Chicago, IL 60637, USA}
\email{ntakoudes@uchicago.edu}

\author[0000-0003-0861-0922]{Hanjue Zhu (朱涵珏)}
\affiliation{
Institute for Advanced Study, 1 Einstein Drive, Princeton, NJ 08540, USA}
\affiliation{Department of Astronomy \& Astrophysics; 
The University of Chicago; Chicago, IL 60637, USA}
\email[show]{hanjuezhu@ias.edu}

\author[0000-0001-5925-4580]{Nickolay Y.\ Gnedin}
\affiliation{Theory Division; 
Fermi National Accelerator Laboratory; Batavia, IL 60510, USA}
\affiliation{Kavli Institute for Cosmological Physics;
The University of Chicago; Chicago, IL 60637, USA}
\affiliation{Department of Astronomy \& Astrophysics; 
The University of Chicago; Chicago, IL 60637, USA}
\email{ngnedin@gmail.com}

\begin{abstract}
Spatial fluctuations in the Thomson optical depth encode information about the inhomogeneous nature of cosmic reionization. We compute the optical-depth angular power spectrum, $C_\ell^{\tau\tau}$, using past lightcones constructed from five Cosmic Reionization on Computers (CROC) radiation-hydrodynamical simulations. By decomposing the electron-density field into patchy and density components, we quantify the separate contributions of ionization-fraction and baryon-density fluctuations to the optical-depth anisotropy. Because the simulations end at $z\approx5$, we supplement the reionization-era signal with an analytic estimate of the fully ionized low-redshift contribution. We find that baryon-density fluctuations dominate the high-redshift signal over most angular scales, while the accumulated low-redshift contribution exceeds the high-redshift signal across the full multipole range considered. Our results demonstrate that a significant fraction of the optical-depth power is not uniquely associated with reionization morphology, implying that future interpretations of $C_\ell^{\tau\tau}$ must account for the density contribution in addition to patchy ionization.
\end{abstract}

\section{Introduction}

Cosmic reionization is one of the last major phase transitions in the history of the Universe, transforming the intergalactic medium (IGM) from a nearly neutral gas into the highly ionized plasma we observe today \citep{Loeb2001,Furlanetto2006,McQuinn2016,Robertson2022}. After recombination, baryons were mostly neutral, and the IGM was opaque to ionizing radiation. As stars, galaxies, and quasars formed, their ultraviolet photons ionized the surrounding gas, carving out expanding \ion{H}{2} regions that grew and eventually filled the IGM. The process was largely complete by $z\approx 6$ \citep{Fan2006,McGreer2015}, but the timing, duration, and topology of reionization remain open questions.

The free electrons produced during reionization leave a distinct imprint on the cosmic microwave background (CMB) through Thomson scattering. CMB photons move freely through space after recombination, but when reionization frees new electrons, a fraction of those photons are scattered out of their original lines of sight. The cumulative probability of scattering along a direction $\hat{\mathbf n}$ is the Thomson optical depth,
\begin{equation}
\tau(\hat{\mathbf n})
=
\sigma_{\rm T}
\int d\chi\,
a(\chi)\,
n_e(\chi\hat{\mathbf n},\chi),
\label{eq:tau}
\end{equation}
where $\sigma_{\rm T}$ is the Thomson cross section, $\chi$ is comoving distance, $a$ is the scale factor, and $n_e$ is the physical free-electron number density. The sky-averaged value $\bar\tau$ suppresses the power spectra of the primary temperature and polarization anisotropies by approximately $e^{-2\bar\tau}$ on scales smaller than the horizon at reionization \citep{Hu2000,Hu2002}. Thomson scattering during reionization also generates a characteristic large-scale ``reionization bump'' in the $E$-mode polarization power spectrum at $\ell\lesssim 10$ \citep{Zaldarriaga1997,Hu2000}. Measurements of $\bar\tau$ therefore constrain the integrated ionization history and help recover the primordial scalar amplitude $A_s$, since small-scale CMB temperature measurements primarily constrain the combination $A_s e^{-2\bar\tau}$ \citep{Planck2018}.

While $\bar\tau$ characterizes the mean scattering history, fluctuations in $\tau(\hat{\mathbf n})$ carry information about the morphology of reionization \citep{Hu2000,Dvorkin2009,Roy2018,Smith2026}. Ionizing sources form preferentially in overdense environments, so reionization is expected to be inhomogeneous. Ionized regions grow outward around galaxies and merge, producing a spatially varying ionization field \citep{Furlanetto2004,McQuinn2007,Zahn2011,Gnedin2022}. Lines of sight through early ionized overdensities accumulate more free electrons than lines of sight through later-ionized or underdense regions. The angular fluctuation field $\delta\tau(\hat{\mathbf n})$ therefore contains information beyond the mean optical depth alone.

The physical origin of this anisotropy is the electron-density field itself. Since
\begin{equation}
n_e(\mathbf{x},z)
=
x_e(\mathbf{x},z)\,
n_b(\mathbf{x},z),
\label{eq:electron_density}
\end{equation}
the optical-depth fluctuations depend on both spatial variations in the electron-per-baryon field $x_e$ and spatial variations in the baryon density $n_b$. The first contribution is directly associated with patchy reionization: it reflects the size, topology, and redshift evolution of ionized regions. The second contribution is present even for a spatially uniform ionization state, because the ionized IGM still traces the underlying density field. Separating these two sources is therefore necessary for identifying which part of the angular power spectrum of the optical depth fluctuations, $C_\ell^{\tau\tau}$, is uniquely tied to reionization morphology.

This distinction has appeared in several related contexts. Early simulation work on secondary CMB anisotropies from reionization showed that the kinetic Sunyaev--Zel'dovich signal is strongly affected by the nonlinear density and velocity fields, rather than being determined only by the detailed distribution of the ionization fraction \citep{Gnedin2001}. Analytic treatments of patchy reionization often model optical-depth fluctuations by projecting the ionization-fraction power spectrum and forecasting its detectability with CMB polarization experiments \citep{Dvorkin2009,Roy2018}. Recent radiation-hydrodynamical lightcone calculations have also computed optical-depth fluctuations directly from simulated reionization histories \citep{Smith2026}. These studies provide the broader context for a direct decomposition of the optical-depth field into its ionization and density contributions.

In this work, we use the Cosmic Reionization on Computers (CROC) radiation-hydrodynamical simulations \citep{Gnedin2014,Gnedin2014b} to compute the angular power spectrum $C_\ell^{\tau\tau}$ of the Thomson optical-depth field. We construct past lightcones from five CROC volumes spanning different large-scale environments and project the simulated electron density to obtain optical-depth maps. We then decompose the electron-density fluctuation into a patchy component, sourced by spatial variations in $x_e$, and a density component, sourced by spatial variations in $n_b$ at fixed mean ionization state. This decomposition allows us to quantify how much of the optical-depth power spectrum is associated with reionization morphology and how much arises from the density field that modulates the electron distribution.

\section{Methodology}
\subsection{Reionization simulations}
\label{sec:sim}
We use the CROC simulations, a suite of cosmological radiation-hydrodynamical simulations designed to follow galaxy formation and intergalactic reionization in representative cosmological volumes \citep{Gnedin2014, Gnedin2014b}. The simulations evolve dark matter, gas dynamics, star formation, feedback, and radiative transfer, and therefore provide the ionization and density fields needed to compute the Thomson optical depth directly from the electron distribution.

For this analysis, we use uniform-grid outputs of the CROC fields in periodic boxes of side length $L_{\rm box}=80\,h^{-1}\,{\rm cMpc}$, with $1024^3$ cells. Each snapshot provides the ionization state, baryon density, and related gas properties on the same grid. 

The simulation set contains five volumes. Three runs, A, B, and C, are three independent realizations of Gaussian initial conditions. Two additional runs are variants of the C realization with manually enhanced initial fluctuations at the box scale -- the so-called ``DC mode" \citep{Pen1997,Sirko2005,Gnedin2011}. The run C.DC$=-1$ samples an underdense large-scale environment, while the run C.DC$=+1$ samples an overdense large-scale environment. The DC mode changes the background density and hence the large scale structure evolution in the simulation, shifting the timing and morphology of reionization. The five boxes therefore sample both realization-to-realization scatter and the response of the optical-depth signal to the large-scale environment.

\begin{figure}[th]
    \includegraphics[width=1\linewidth]{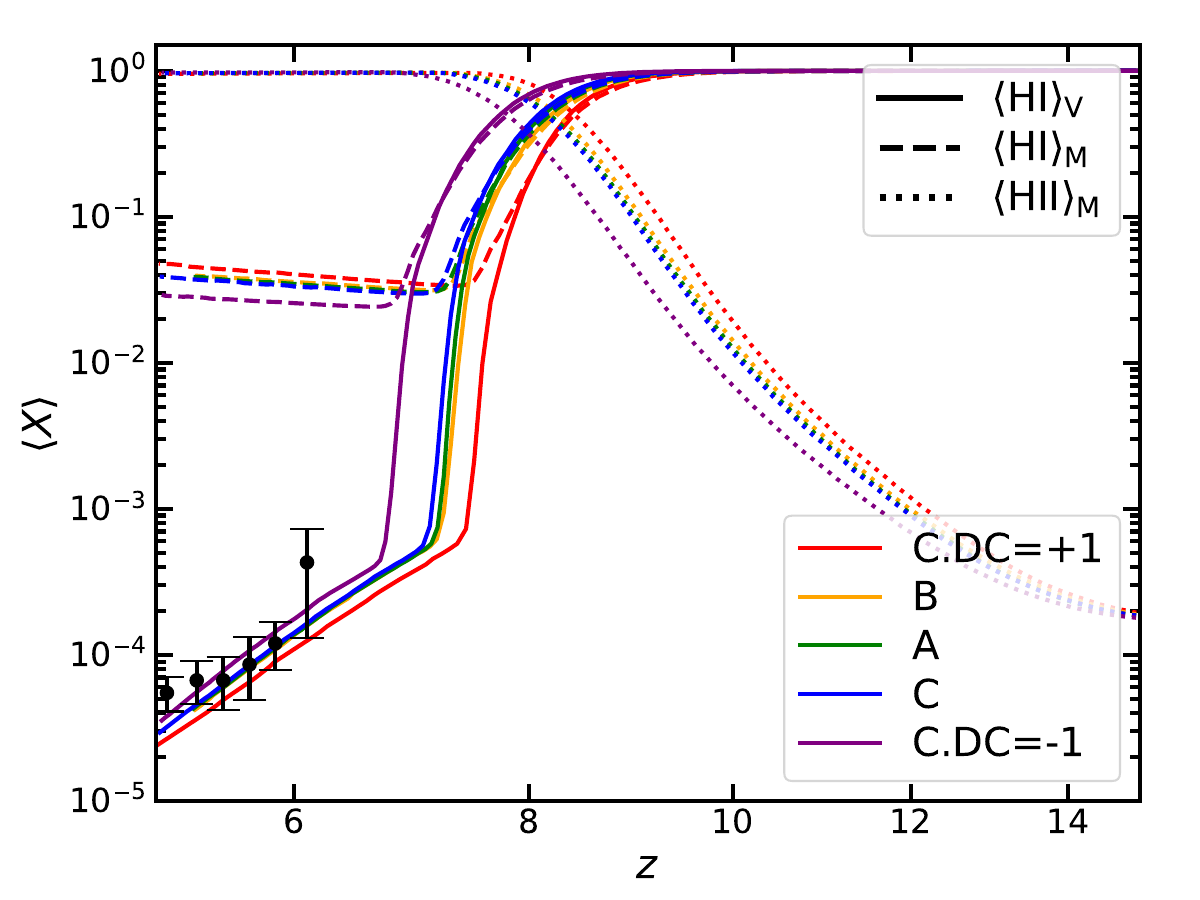}
    \caption{Reionization histories for all five CROC simulations. The neutral hydrogen fraction $\langle \mathrm{HI} \rangle$ is shown as both volume-weighted (solid lines) and mass-weighted (dashed lines) averages as a function of redshift. Ionized hydrogen fraction $\langle \mathrm{HII} \rangle$ is shown as mass-weighted only (dotted lines) - the latter is the quantity used below in Equation (\ref{eq:xebar}). The different colors correspond to the different CROC boxes: A (green), B (orange), C (blue), C.DC$=+1$ (red), C.DC$=-1$ (purple), in the rainbow order from the earliest to the latest reionization history. Black points show simulation calibration data from \cite{Fan2006}.\label{fig:reionization_histories}}
\end{figure}

Figure~\ref{fig:reionization_histories} shows the reionization histories of the five boxes. The spread among these histories is part of the physical variation propagated into the optical-depth power spectra.

\subsection{From simulation snapshots to optical-depth maps}
\label{sec:tau_map}
The optical-depth field is a line-of-sight projection of the physical free-electron density (Equation \ref{eq:tau}), which we evaluate directly from the CROC simulation snapshots by constructing a discrete past lightcone. For a snapshot at scale factor $a$, the corresponding comoving radial distance is
\begin{equation}
    \chi(a) = \int_a^1 \frac{c\,da'}{a'^2H(a')},
\end{equation}
with
\begin{equation}
    H(a) = H_0\sqrt{\Omega_m a^{-3}+\Omega_\Lambda},
\end{equation}
and cosmological parameters $h=0.68$, $\Omega_b=0.0479$, $\Omega_m=0.3036$, and $\Omega_\Lambda=0.6964$. In the flat-sky description below, $\hat{\mathbf n}(\boldsymbol{\theta})$ denotes the line-of-sight unit vector associated with angular coordinate $\boldsymbol{\theta}$.

The output of a CROC simulation is a sequence of discrete snapshots, ordered by increasing scale factor. There are two alternative ways to construct a lightcone from these data. In the first approach, which we call ``tiles,'' we tile the radial axis of the lightcone with full simulation boxes ($80\,{\rm cMpc}/h$). The centers of these tiled boxes do not generally coincide with the times of the simulation snapshots. We therefore linearly interpolate the snapshot data in time to the center of each tiled box.

In the second approach, which we call ``slabs,'' we center simulation boxes along the lightcone at the exact times of the simulation snapshots. The distance between neighboring snapshots is always smaller than one box length ($80\,{\rm cMpc}/h$). To ensure non-overlapping radial coverage, we therefore select a random portion of each simulation box, with the depth of each portion chosen so that these ``slabs'' fully cover the radial axis without overlap.

Both constructions reuse the same finite periodic volume along the line of sight. We therefore apply independent random transformations to successive tiles or slabs to reduce artificial coherence between repeated images. Each contribution is assigned a random projection axis, a random rotation by a multiple of $90^\circ$, random periodic translations in the two transverse directions, and independent reflections about the transverse axes. These operations preserve the statistical properties of the snapshot while changing the relative positions and orientations of structures in neighboring lightcone elements, following the general randomization strategy of \citet{Gnedin2001}.

The optical depth in each map pixel is accumulated along one fixed sky direction $\boldsymbol{\theta}$. At comoving distance $\chi_i$, that direction intersects transverse coordinates $\boldsymbol{x}_{\perp}=\chi_i\boldsymbol{\theta}$, so the same angular field covers a different comoving region at each redshift. We therefore place all projected maps on a common sky grid by choosing a reference scale factor $a_{\rm ref}$, which defines a reference comoving distance $\chi_{\rm ref}$ and a common angular pixel size
$$
\Delta\theta_{\rm ref} =
\frac{\Delta x_{\rm com}}{\chi_{\rm ref}},
$$
where $\Delta x_{\rm com}=L_{\rm box}/N_{\rm pix}$ is the comoving pixel size of a projected map with $N_{\rm pix}$ pixels on a side. At distance $\chi_i$, this fixed angular grid subtends a transverse comoving region of size
$$
L_{\perp,i} =
L_{\rm box}\frac{\chi_i}{\chi_{\rm ref}}.
$$
At $a<a_{\rm ref}$, a simulation box subtends a smaller angular region
than the fixed sky field, so periodic replications of the box are used
to fill the field. At $a>a_{\rm ref}$, a simulation box subtends a
larger angular region than the fixed sky field, so only part of the box
is used to make the sky map.

We denote the resulting optical-depth fluctuation contribution from lightcone element $i$ (a tile or a slab), including the time interpolation for tiles and the angular resampling, by $\Delta\delta\tau_i^{\rm interp}(\boldsymbol{\theta})$. The final simulated optical-depth fluctuation map is
\begin{equation}
\delta\tau(\boldsymbol{\theta}) =
\sum_i \Delta\delta\tau_i^{\rm interp}(\boldsymbol{\theta}).
\end{equation}

\begin{figure}[th]
    \centering
    \includegraphics[width=\columnwidth]{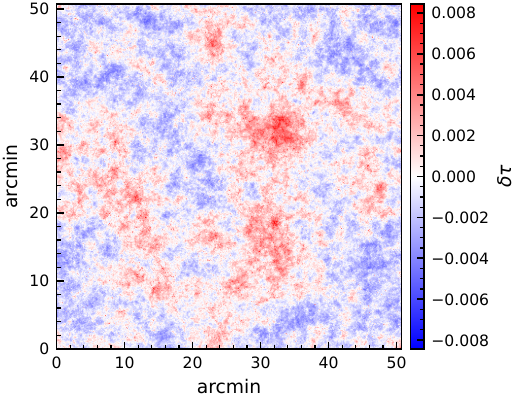}
    \caption{Map of the mean-subtracted optical-depth fluctuation
    $\delta\tau(\boldsymbol{\theta})$ for realization C, constructed from the
    lightcone with $a_{\rm ref}=0.1667$ and summing the patchy and density
    contributions. The color scale spans $\pm5\sigma$ about the mean, where
    $\sigma$ is the standard deviation of $\delta\tau$ across the map.}
    \label{fig:tau_map}
\end{figure}

This procedure correctly accounts for the redshift dependence of angular diameter distance before the contributions are summed, and is the same as used by the Lumina project \citep{Smith2026}. The justification for this procedure is that the power spectrum is largely independent of $a_{\rm ref}$, as shown in Appendix \ref{appendix:reference}. Before taking the Fourier transform, we subtract the angular mean of each final component map, thereby removing the $\ell=0$ mode. An example of $\delta\tau(\boldsymbol{\theta})$ is shown in Figure~\ref{fig:tau_map} for the default ``tile'' lightcone construction.

\subsection{Angular power-spectrum estimation}

For all reference scale factors considered here, the lightcone field has angular side length $\Theta_{\rm map}=L_{\rm box}/\chi_{\rm ref}\lesssim0.85^\circ$. Sky curvature across the field is therefore negligible, and we use the flat-sky approximation. Directions within the field are represented by tangent-plane angular coordinates $\boldsymbol{\theta}=(\theta_x,\theta_y)$ in radians, with Fourier-conjugate two-dimensional multipole vector $\boldsymbol{\ell}=(\ell_x,\ell_y)$.

For a mean-subtracted optical-depth map $\delta\tau(\boldsymbol{\theta})$, we define the flat-sky Fourier transform
\begin{equation}
    \delta\tau(\boldsymbol{\ell})
    =
    \int d^2\theta\,
    \delta\tau(\boldsymbol{\theta})
    e^{-i\boldsymbol{\ell}\cdot\boldsymbol{\theta}}.
\end{equation}
The angular area of the square field is $A=\Theta_{\rm map}^2$, and its discrete Fourier modes are
\begin{equation}
    \ell_x=\frac{2\pi m}{\Theta_{\rm map}},
    \qquad
    \ell_y=\frac{2\pi n}{\Theta_{\rm map}},
\end{equation}
where $m$ and $n$ are integer FFT indices. The finite angular pixel size also sets the maximum reliably sampled
multipole. For the common grid spacing $\Delta\theta_{\rm ref}$, the
one-dimensional Nyquist multipole is
\begin{equation}
    \ell_{\rm Ny}
    =
    \frac{\pi}{\Delta\theta_{\rm ref}}.
\end{equation}
Although the square two-dimensional FFT grid contains corner modes with
$|\boldsymbol{\ell}|>\ell_{\rm Ny}$, these modes do not form complete
isotropic annuli. We therefore restrict the azimuthally binned power
spectrum to
\begin{equation}
    \ell \leq \ell_{\rm Ny}.
\end{equation}

We evaluate the discrete transform as
\begin{equation}
    \delta\tau(\boldsymbol{\ell}_{mn})
    =
    \Delta\theta_{\rm ref}^2
    \sum_{j,k}
    \delta\tau(\boldsymbol{\theta}_{jk})
    e^{-i\boldsymbol{\ell}_{mn}\cdot\boldsymbol{\theta}_{jk}},
\end{equation}
so that the discrete sum approximates the continuum transform as a Riemann sum.

For an isotropic field, the angular power spectrum is defined by
\begin{equation}
    \left\langle
    \delta\tau(\boldsymbol{\ell})
    \delta\tau^*(\boldsymbol{\ell}')
    \right\rangle
    =
    (2\pi)^2
    \delta^{(2)}(\boldsymbol{\ell}-\boldsymbol{\ell}')
    C_\ell^{\tau\tau}.
\end{equation}
For a finite map, using $(2\pi)^2\delta^{(2)}(\boldsymbol{0})\rightarrow A$, this becomes
\begin{equation}
    \left\langle
    \left|
    \delta\tau(\boldsymbol{\ell})
    \right|^2
    \right\rangle
    =
    A\,C_\ell^{\tau\tau}.
\end{equation}
The practical binned estimator is then
\begin{equation}
    \widehat C_b^{\tau\tau}
    =
    \frac{1}{A}
    \frac{1}{N_b}
    \sum_{\boldsymbol{\ell}\in b}
    \left|
    \delta\tau(\boldsymbol{\ell})
    \right|^2,
\end{equation}
where $N_b$ is the number of discrete Fourier modes in multipole bin $b$. The measured spectrum therefore describes fluctuations in optical depth rather than the sky-averaged optical depth.

\subsection{Patchy and density contributions to optical-depth fluctuations}

We decompose the simulated optical-depth fluctuation into two physically distinct contributions: one sourced by spatial variations in the ionization state and one sourced by spatial variations in the baryon density. The decomposition is defined separately at each simulation snapshot, before temporal interpolation and angular resampling.

The electron density is related to the baryon density and electron-per-baryon field through Eq.~\eqref{eq:electron_density}. Here $n_b$ is the physical baryon number density and $x_e\equiv n_e/n_b$ is the number of free electrons per baryon. For each snapshot, we define
\begin{equation}
\bar n_b\equiv \langle n_b\rangle,
\qquad
\bar n_e\equiv \langle n_e\rangle,
\end{equation}
where the averages are taken over the full simulation volume, and
\begin{equation}
\bar x_e
\equiv
\frac{\bar n_e}{\bar n_b}.
\label{eq:xebar}
\end{equation}

At a fixed cosmic time, the electron-density fluctuation is
\begin{align}
\delta n_e
&\equiv
n_e-\bar n_e
\nonumber\\
&=
x_e n_b-\bar x_e\bar n_b
\nonumber\\
&=
(x_e-\bar x_e)n_b
+
\bar x_e(n_b-\bar n_b).
\end{align}
This identity defines the patchy contribution,
\begin{equation}
\delta n_{e,\mathrm{patchy}}
\equiv
(x_e-\bar x_e)n_b,
\end{equation}
and the density contribution,
\begin{equation}
\delta n_{e,\mathrm{density}}
\equiv
\bar x_e(n_b-\bar n_b).
\end{equation}
The patchy term vanishes when the ionization fraction is spatially uniform and therefore traces spatial variations in the ionization state. The density term is the electron-density fluctuation that would remain for the same mean ionization state in a spatially fluctuating baryon field. By construction,
\begin{equation}
\delta n_e
=
\delta n_{e,\mathrm{patchy}}
+
\delta n_{e,\mathrm{density}}.
\label{eq:patchy_density}
\end{equation}

The corresponding optical-depth fluctuation components are
\begin{equation}
\delta\tau_{\mathrm{patchy}}(\hat{\mathbf n})
=
\sigma_T
\int
\delta n_{e,\mathrm{patchy}}(\chi\hat{\mathbf n},\chi)\,
dl_{\rm proper},
\end{equation}
and
\begin{equation}
\delta\tau_{\mathrm{density}}(\hat{\mathbf n})
=
\sigma_T
\int
\delta n_{e,\mathrm{density}}(\chi\hat{\mathbf n},\chi)\,
dl_{\rm proper}.
\end{equation}
Their sum gives the simulated optical-depth fluctuation,
\begin{equation}
\delta\tau(\hat{\mathbf n})
=
\delta\tau_{\mathrm{patchy}}(\hat{\mathbf n})
+
\delta\tau_{\mathrm{density}}(\hat{\mathbf n}).
\end{equation}

We apply the flat-sky estimator defined above to the patchy and density maps after subtracting their angular means. For a multipole bin $b$ containing $N_b$ discrete Fourier modes and a map of angular area $A$, the binned auto-spectra are
\begin{equation}
\widehat C_b^{\mathrm{patchy}}
=
\frac{1}{A}
\frac{1}{N_b}
\sum_{\boldsymbol{\ell}\in b}
\left|
\delta\tau_{\mathrm{patchy}}(\boldsymbol{\ell})
\right|^2,
\end{equation}
and
\begin{equation}
\widehat C_b^{\mathrm{density}}
=
\frac{1}{A}
\frac{1}{N_b}
\sum_{\boldsymbol{\ell}\in b}
\left|
\delta\tau_{\mathrm{density}}(\boldsymbol{\ell})
\right|^2.
\end{equation}
The corresponding patchy--density cross spectrum is
\begin{equation}
\widehat C_b^{\mathrm{patchy}\times\mathrm{density}}
=
\frac{1}{A}
\frac{1}{N_b}
\sum_{\boldsymbol{\ell}\in b}
\mathrm{Re}\left[
\delta\tau_{\mathrm{patchy}}(\boldsymbol{\ell})
\delta\tau_{\mathrm{density}}^{*}(\boldsymbol{\ell})
\right].
\end{equation}
Since the total fluctuation map is the sum of the two component maps, the spectra satisfy
\begin{equation}
\widehat C_b^{\tau\tau}
=
\widehat C_b^{\mathrm{patchy}}
+
\widehat C_b^{\mathrm{density}}
+
2\,\widehat C_b^{\mathrm{patchy}\times\mathrm{density}}.
\label{eq:patchy_density_decomp}
\end{equation}
We use this relation as a consistency check on the map construction and Fourier normalization.

\subsection{Low-redshift contribution}

The CROC lightcones used in this work terminate at the lowest available simulation output, corresponding to approximately $z\simeq 5$. They therefore cover the reionization-era contribution to the optical-depth fluctuations but do not include the fully ionized, lower-redshift universe. At lower redshift, after hydrogen reionization, we assume a spatially uniform ionization state, while baryon-density fluctuations still generate optical-depth anisotropy. We therefore add the low-redshift density contribution analytically over the interval $0<z<z_{\rm cut}$, where $z_{\rm cut}$ is the lower-redshift boundary of the simulation lightcone.

The proper electron density in this regime is
\begin{equation}
n_e^{\mathrm{proper}}(\mathbf{x},z)
=
f_e(z)\,\bar n_{b,0}\,a^{-3}(z)
\left[1+\delta_b(\mathbf{x},z)\right],
\end{equation}
where $\bar n_{b,0}$ is the present-day mean baryon number density and $f_e(z)$ is the number of free electrons per baryon. When helium is singly ionized, the number of free electrons per baryon is
\begin{equation}
f_e = X_{\rm H}+\frac{Y_{\rm He}}{4}=0.82,
\end{equation}
while after helium double ionization it is
\begin{equation}
f_e = X_{\rm H}+\frac{Y_{\rm He}}{2}=0.88.
\end{equation}
We assume helium becomes doubly ionized at $z=3$, and therefore use $f_e=0.82$ for $z>3$ and $f_e=0.88$ for $z\le3$. 

The mean part of $n_e^{\rm proper}$ contributes only to the sky-averaged optical depth. The fluctuating low-redshift contribution is
\begin{equation}
\delta\tau_{\mathrm{low}\text{-}z}(\hat{\mathbf n})
=
\sigma_T \bar n_{b,0}
\int_0^{\chi_{\mathrm{cut}}} d\chi\,
\frac{f_e(z)}{a^2(\chi)}\,
\delta_b(\chi\hat{\mathbf n},z).
\end{equation}
The factor $a^{-2}$ comes from combining the proper electron-density scaling, $n_e^{\rm proper}\propto a^{-3}$, with the proper line element, $dl_{\rm proper}=a\,d\chi$. This expression has the form of a line-of-sight projection of the three-dimensional density field. In the Limber approximation, an angular mode $\ell$ mainly probes transverse comoving wavenumber $k\simeq(\ell+1/2)/\chi$ at distance $\chi$. Applying this approximation gives
\begin{equation}
\begin{aligned}
C_\ell^{\tau\tau,\mathrm{low}\text{-}z}
\simeq\;
& \sigma_T^2 \bar n_{b,0}^2
\int_{z_{\min}(\ell)}^{z_{\mathrm{cut}}} dz\,
\frac{c}{H(z)} \\
& \times
\frac{f_e^2(z)}{a^4(z)\chi^2(z)}
\,P_{\delta\delta}\!\left(
k=\frac{\ell+1/2}{\chi(z)},z
\right),
\end{aligned}
\end{equation}
where we have approximated the baryon-density power spectrum by the total matter power spectrum, $P_{\delta_b\delta_b}\simeq P_{\delta\delta}$. We compute $P_{\delta\delta}(k,z)$ using the nonlinear matter power
spectrum from \textsc{CAMB}. The spectrum is calculated up to
$k_{\rm CAMB,max}=1000\,{\rm Mpc}^{-1}$, and the matter-power
interpolator is allowed to extrapolate to
$k_{\rm ext,max}=5000\,{\rm Mpc}^{-1}$.

For each multipole, we impose a lower redshift limit so that the Limber
wavenumber does not exceed the maximum extrapolated wavenumber,
\begin{equation}
\chi[z_{\min}(\ell)]
=
\frac{\ell+1/2}{k_{\rm ext,max}}.
\end{equation}
The interval $k_{\rm CAMB,max}<k\le k_{\rm ext,max}$ therefore uses the
extrapolated matter power spectrum.

The full optical-depth fluctuation is the sum of the simulated reionization-era map and the low-redshift density contribution,
\begin{equation}
\delta\tau(\hat{\mathbf n})
=
\delta\tau_{\mathrm{sim}}(\hat{\mathbf n})
+
\delta\tau_{\mathrm{low}\text{-}z}(\hat{\mathbf n}) .
\end{equation}
The exact angular power spectrum contains the two auto-spectra and their cross spectrum,
\begin{equation}
C_\ell^{\tau\tau}
=
C_\ell^{\tau\tau,\mathrm{sim}}
+
C_\ell^{\tau\tau,\mathrm{low}\text{-}z}
+
2C_\ell^{\mathrm{sim}\times\mathrm{low}\text{-}z}.
\end{equation}
In the Limber approximation, the cross spectrum vanishes for non-overlapping radial intervals. We therefore neglect the cross term and use
\begin{equation}
C_\ell^{\tau\tau}
\approx
C_\ell^{\tau\tau,\mathrm{sim}}
+
C_\ell^{\tau\tau,\mathrm{low}\text{-}z}.
\end{equation}

\section{Results}

\begin{figure*}[t!]
\centering
\includegraphics[width=\textwidth]{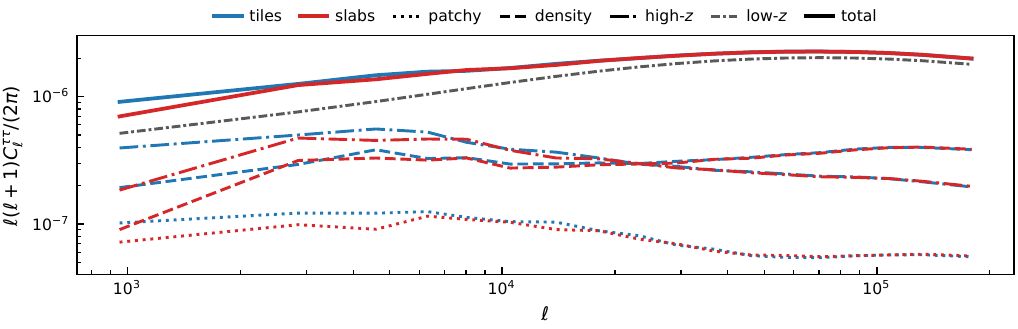}
\caption{Angular power spectrum
$\ell(\ell+1)C_\ell^{\tau\tau}/2\pi$ for realization C using
$a_{\rm ref}=0.1667$. The blue curves show the ``tiles''
lightcone construction, including the patchy
(dotted), density (dashed), high-$z$ total (dash-dotted), and complete
total (solid) contributions. The red curves show the alternative
``slabs'' lightcone construction, with the same high-$z$ decomposition
into patchy (dotted), density (dashed), and high-$z$ total (dash-dotted)
components. The shared low-$z$ density contribution is shown once as the gray dash-dotted curve. The solid red curve shows the
complete slab result obtained by adding this same low-$z$ contribution
to the slab high-$z$ spectrum. The difference between the tile and slab
constructions provides a measure of the systematic uncertainty associated
with the lightcone construction. The contribution
$2\widehat{C}_\ell^{\mathrm{patchy}\times\mathrm{density}}$ is omitted
because it is negative at high $\ell$, making it unsuitable for display
on a logarithmic scale. The total signal is dominated by the low-$z$
density contribution, while the high-$z$ signal is itself density
dominated.}
\label{fig:c_spectrum}
\end{figure*}

Figure~\ref{fig:c_spectrum} shows a representative decomposition of
$\ell(\ell+1)C_\ell^{\tau\tau}/2\pi$ for CROC realization C. We show our
results for both the ``tiles'' and ``slabs'' lightcone construction
procedures. Since there is no \textit{a priori} reason to choose one
procedure over the other, the difference between the two serves as a measure of systematic sensitivity to our lightcone construction.

The prominent feature of Figure~\ref{fig:c_spectrum} is that the simulated
high-redshift signal is already density dominated. The density component
exceeds the patchy component at every multipole shown, typically by a
factor of a few. Thus, even during the reionization era, the optical-depth power is set primarily by fluctuations in the cosmic gas density rather than by spatial variations in the ionization state.

The decomposition of Equation~\eqref{eq:patchy_density_decomp} also
constrains the sign of the patchy--density cross term. At
$\ell\gtrsim2\times10^4$, the high-redshift total falls below the density
component alone, which requires
\begin{equation}
2C_\ell^{\rm patchy\times density}
<
-C_\ell^{\rm patchy}.
\end{equation}
The cross term $2C_\ell^{\rm patchy\times density}$ is therefore negative on these scales, with a magnitude
exceeding the patchy contribution itself. A plausible physical explanation
of this behavior is that small-scale overdense gas has enhanced
recombination rates and can be more strongly self-shielded, leaving its
ionized fraction lower than that of its surroundings. In that case,
positive density fluctuations can correlate with negative fluctuations in
the local electron-per-baryon field.

\begin{figure*}[htb!]
\centering
\includegraphics[width=1\linewidth]{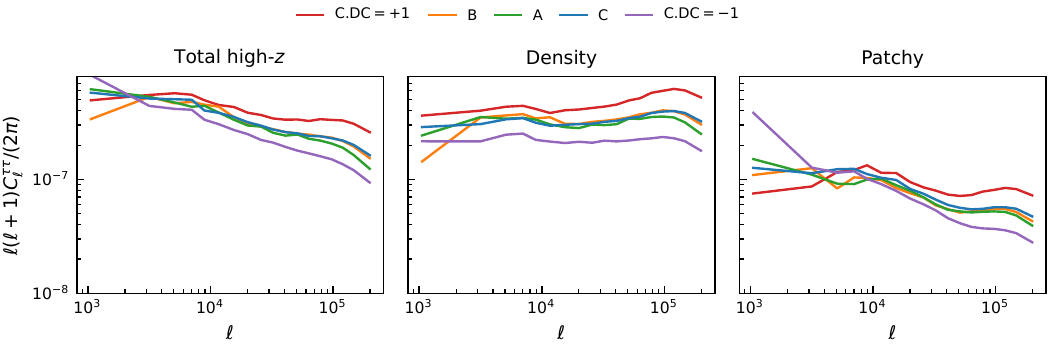}
\caption{High-redshift ($z\gtrsim5$) optical-depth power spectra,
$\ell(\ell+1)C_\ell^{\tau\tau}/2\pi$, from the ``tiles'' lightcone
construction using $a_{\rm ref}=0.1667$. The left, middle, and right
panels show the total, density, and patchy contributions, respectively.
Colors identify the CROC realizations: A (green), B (orange), C (blue),
and the positive- and negative-DC-mode variants of one realization
(red and purple, respectively). Across all five simulations, the
high-redshift signal is predominantly density sourced, while the patchy
component shows a scale-dependent response to the DC mode.}
\label{fig:all_highz}
\end{figure*}

Figure~\ref{fig:all_highz} extends the comparison to all five simulations.
The three independent realizations A, B, and C show similar total,
density, and patchy spectra over most of the multipole range, with larger
variation toward the lowest multipoles, where the finite simulation volume
contains few independent large-scale modes. In the density-sourced
spectrum, the DC-mode variants remain clearly separated across the full
multipole range, as expected from their different mean densities, with the
separation becoming larger toward high $\ell$. The total spectra show the
same ordering at high $\ell$.

The patchy contribution instead reverses its DC-mode ordering with scale.
C.DC=$+1$ has the largest patchy power at high $\ell$ and the smallest at
the lowest multipoles, where C.DC=$-1$ has the largest of the five. Later
reionization in the underdense volume may preserve large-scale structure
in the ionization field over a longer portion of the lightcone, while the
greater abundance of collapsed structure and sources in the overdense
volume can imprint more small-scale structure on the ionization field.
Later reionization also places the patchy signal of C.DC=$-1$ at smaller
comoving distance, shifting a fixed comoving scale toward lower $\ell$.
These spectra do not separate this geometric effect from changes in the
morphology of the ionization field.

The analytic low-redshift contribution is larger than any contribution
from reionization. Below $z_{\rm cut}$, the ionization state is nearly
uniform, so this term is density sourced to the accuracy of our
calculation. Its amplitude reflects the long comoving path length from
$z=0$ to $z_{\rm cut}$ and the growth of nonlinear structure at late
times. Once included, it dominates the total spectrum over most of the
plotted range. Measurements or forecasts of the full optical-depth power
spectrum therefore cannot be interpreted as measurements of patchy
reionization morphology without explicitly accounting for the
density-sourced contribution.

\section{Conclusions}
We have used the CROC simulations to compute the angular power spectrum of the Thomson optical-depth field imprinted on the CMB by cosmic reionization. By constructing past lightcones from five CROC simulation volumes, including three independent realizations and two DC-mode variants of one realization, and projecting the simulated electron density, we obtain optical-depth maps. We measure their power spectrum $C_\ell^{\tau\tau}$ directly from the simulated ionization and density fields. Because the CROC lightcones terminate at $z\simeq 5$, we supplement the simulated reionization-era signal with an analytic estimate of the fully ionized, lower-redshift contribution, computed from the nonlinear matter power spectrum in the Limber approximation, so that the combined result spans the full path length from $z=0$ to the start of reionization.

We decompose the electron-density fluctuation, and therefore $\delta\tau(\hat{\mathbf n})$, into a patchy component sourced by spatial variations in the ionization fraction $x_e$ and a density component sourced by spatial variations in the baryon density $n_b$ at fixed mean ionization state. This split is possible because we work directly with the simulated electron-density field, and it isolates the part of $C_\ell^{\tau\tau}$ that is genuinely tied to the morphology of reionization from the part that would be present even if reionization proceeded homogeneously.

We find that the density component dominates the high-redshift ($z\gtrsim5$) contribution to $C_\ell^{\tau\tau}$ over nearly the full multipole range in all five simulations, with the exception of the
underdense DC-mode realization at the lowest multipoles. The patchy contribution is therefore subdominant even during the partially ionized epoch. The DC-mode variants also reveal a scale-dependent response of the patchy signal: the underdense volume has the largest patchy power on the largest scales, while the overdense volume has the largest patchy power at high $\ell$. At sufficiently high multipoles, the total high-redshift power falls below the density contribution alone in every simulation, requiring a negative patchy--density cross term. Once the analytic low-redshift contribution is included, the density-sourced low-redshift signal dominates the total optical-depth power over the whole plotted range, reflecting both the long comoving path length through the fully ionized universe and the growth of nonlinear structure at late times.

These results bear directly on efforts to extract reionization morphology from measurements or forecasts of $C_\ell^{\tau\tau}$: because the density-sourced component is not itself diagnostic of patchy reionization, isolating the reionization-morphology signal from the total optical-depth power spectrum requires either modeling and subtracting the density contribution or adopting a statistic that suppresses it, rather than treating the full $C_\ell^{\tau\tau}$ as tracing patchy reionization directly. Analytic forecasts that model $\delta\tau$ through the projected ionization-fraction power spectrum alone should therefore account for the density contribution quantified here, particularly at the multipoles where it dominates.

The present analysis is limited by the use of five simulation volumes, which constrains our ability to quantify volume-to-volume and large-scale-environment (DC-mode) sample variance in $C_\ell^{\tau\tau}$, and by treating the low-redshift, fully ionized contribution analytically rather than from simulations extending to $z=0$. Future work should extend the CROC lightcone approach to a larger simulation ensemble in order to bound the environmental scatter in the patchy and density power spectra directly, resolve the origin of the negative patchy--density cross-correlation identified here, and connect this decomposition to forecasts for CMB observables sourced by optical-depth fluctuations.

\appendix
\section{Justification of the Reference Angular Grid}
\label{appendix:reference}

\begin{figure*}[htb!]
    \centering
    \includegraphics[width=\textwidth]{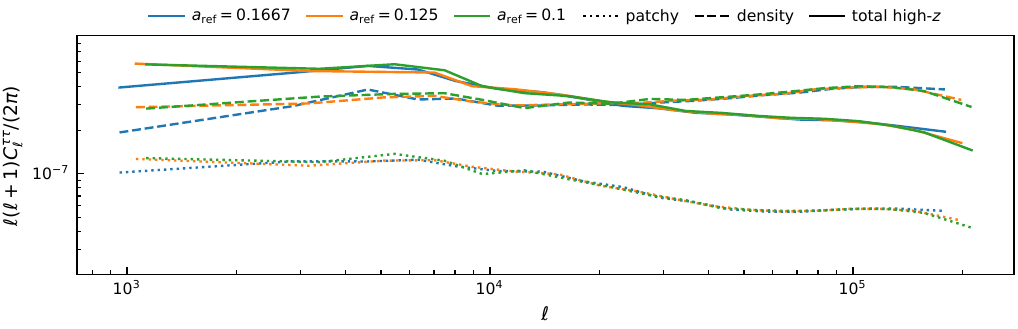}
    \caption{Angular power spectra $\ell(\ell+1)C_\ell^{\tau\tau}/2\pi$ of realization C computed using three different reference scale factors $a_{\rm ref} = 0.1667$ (blue), $0.125$ (orange), and $0.10$ (green), corresponding to different choices of the reference angular grid. The high-$z$ total (solid), density (dashed), and patchy (dotted) components are shown for each choice. The three choices give similar spectra over the central multipole
range, with larger differences toward the lowest and highest
multipoles, while preserving the same qualitative decomposition of
the high-redshift signal.}
    \label{fig:aref}
\end{figure*}

As described in Section~\ref{sec:tau_map}, each projected simulation cube is resampled onto a common angular grid before constructing the final optical-depth map. This requires choosing a reference angular pixel size, which we define by
\begin{equation}
\Delta\theta_{\rm ref}
=
\frac{\Delta x_{\rm com}}{\chi_{\rm ref}},
\end{equation}
where $\chi_{\rm ref}=\chi(a_{\rm ref})$ is the comoving distance corresponding to the chosen reference scale factor $a_{\rm ref}$. For our fiducial choice, $a_{\rm ref}=0.1667$. Changing $a_{\rm ref}$ changes both the angular pixel size and the angular extent of the fixed-$N$ map, so we
test the dependence of the resulting power spectra on this choice.

Figure~\ref{fig:aref} compares the angular power spectra computed using three different reference scale factors, $a_{\rm ref}=0.1667$, $0.125$, and $0.1$, corresponding to different choices of the reference angular resolution. For each choice, the lightcone is reconstructed using identical cubes, random translations, and reflections; only the angular grid onto which the individual cubes are interpolated is changed. The three calculations track one another most closely over the intermediate multipoles. The density and total spectra show more visible differences toward the lowest multipoles, where the finite angular extent of the map leaves only a small number of discrete Fourier modes and changing $a_{\rm ref}$ shifts the corresponding low-$\ell$ mode sampling. Differences also grow at the highest multipoles, where the angular scales approach the grid resolution and the resampling becomes more important.

Across the plotted multipole range, the dependence on the reference grid does not alter the separation between the physical components. For all three choices, the density contribution exceeds the patchy contribution, and the overall shapes and relative ordering of the components are preserved.

\begin{acknowledgments}
The code used to generate all figures and results in this paper is publicly available at \url{https://github.com/hanjuezhu/Optical_Depth_Reionization}. NT acknowledges support from the T.D. Lee Undergraduate Summer Research Fellowship. HZ gratefully acknowledges support from the Institute for Advanced Study. This work was supported in part by Fermi Forward Discovery Group, LLC, under Contract No.\ 9243024CSC000002 with the U.S. Department of Energy, Office of Science, Office of High Energy Physics. This work used resources of the Argonne Leadership Computing Facility, which is a DOE Office of Science User Facility supported under Contract DE-AC02-06CH11357. An award of computer time was provided by the Innovative and Novel Computational Impact on Theory and Experiment (INCITE) program. This research is also part of the Blue Waters sustained-petascale computing project, which is supported by the National Science Foundation (awards OCI-0725070 and ACI-1238993) and the state of Illinois. Blue Waters is a joint effort of the University of Illinois at Urbana-Champaign and its National Center for Supercomputing Applications. 
\end{acknowledgments}

\bibliography{main}{}
\bibliographystyle{aasjournal}

\end{CJK*}
\end{document}